       \documentclass[final,authoryear,5p,times]{elsarticle}

\usepackage{graphicx,amssymb,amsthm,bm,lineno}
\usepackage{apalike}
\usepackage{color}
\newcommand{\executeiffilenewer}[3]{%
\ifnum\pdfstrcmp{\pdffilemoddate{#1}}%
{\pdffilemoddate{#2}}>0%
{\immediate\write18{#3}}\fi%
}

\biboptions{longnamesfirst,comma}

\begin{document}

\begin{frontmatter}



\title{Universal kinetics of the onset of cell spreading on substrates of different stiffness}


\author{Samuel Bell, Anna-Lena Redmann and Eugene M. Terentjev }

\address{Cavendish Laboratory, University of Cambridge, Cambridge CB3 0HE, U.K.}

\begin{abstract}
{When plated onto substrates, cell morphology and even stem cell differentiation are influenced by the stiffness of their environment. Stiffer substrates give strongly spread (eventually – polarized) cells with strong focal adhesions, and stress fibers; very soft substrates give a less developed cytoskeleton, and much lower cell spreading. The kinetics of this process of cell spreading is studied extensively, and important universal relationships are established on how the cell area grows with time. Here we study the population dynamics of spreading cells, investigating the characteristic processes involved in cell response to the substrate. We show that unlike the individual cell morphology, this population dynamics does not depend on the substrate stiffness. Instead, a strong activation temperature dependence is observed. Different cell lines on different substrates all have long-time statistics controlled by the thermal activation over a single energy barrier $\Delta G\approx 19$ kcal/mol, while the early-time kinetics follows a power law $\sim t^5$. This implies that the rate of spreading depends on an internal process of adhesion complex assembly and activation: the operational  complex must have 5 component proteins, and the last process in the sequence (which we believe is the activation of focal adhesion kinase) is controlled by the binding energy $\Delta G$.  }
\end{abstract}


\end{frontmatter}



\section{Introduction}
\label{sec:intro}

Matrix stiffness is known to affect cell size and morphology~\citep{Discher2005,Yeung2005}. When cells are plated onto soft substrates, their footprint will not increase as much as on stiff substrates, and their spreading will be more isotropic: resulting cells will be round and dome-like in shape. On stiff substrates, the same cells will spread very strongly, develop concentrated focal adhesion clusters and stress fibers of bundled F-actin, and eventually polarize to initiate migration. This leads to several well-documented biological functions in tissues: variable stem-cell differentiation pathways~\citep{Discher2005,Engler2006}, the fibroblast-myofibroblast transition near scar tissue~\citep{Hinz2007,Tomasek2002,Solon2007}, fibrosis in smooth-muscle cells near rigid plaque or scar tissue~\citep{Sinha2004,Cheung2012}, and the stiffer nature of tumor cells~\citep{Alliston2001,Butcher2009}. The definitive review \citep{Schwarz-review} summarizes this topic.

\begin{figure}
	\centering
	\includegraphics[width=0.35\textwidth]{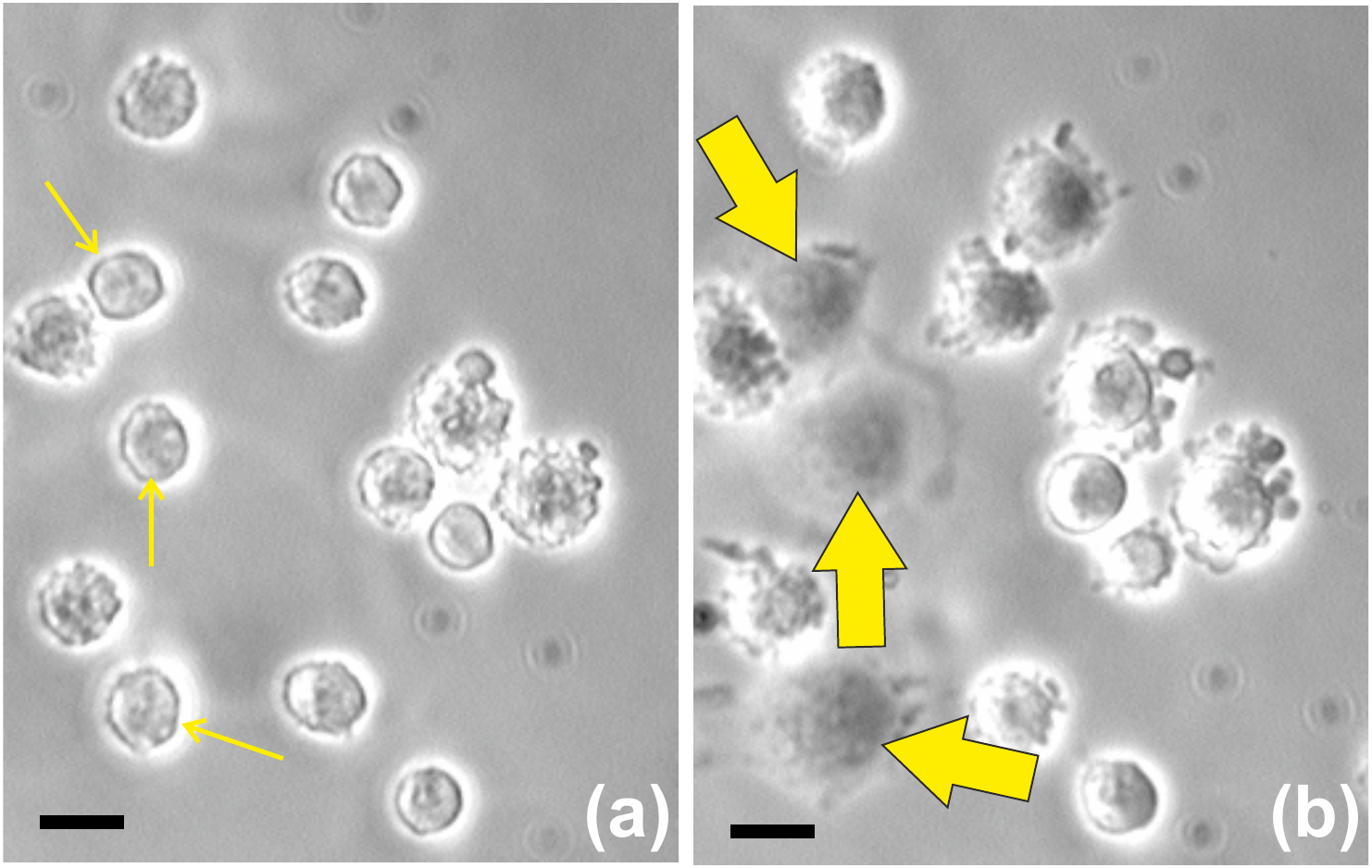}
	\caption{A section of the experimental field of view, illustrating the onset of spreading. Photographs (a) and (b) show the same cells: immediately after planting on the substrate (solid glass with fibronectin), and 15 min later, when several cells have already responded by spreading (labelled by matching arrows). Scale bar = 20 $\mu$m. }
	\label{fig:cells}
\end{figure}

The dynamics of cells spreading has been studied extensively, and several characteristic universal features have been established \citep{Schwarz-review,Mahadevan2007,Krishnan2014,Reinhart2005}. In particular, the average cell area has been shown to grow with time as a power law, often with the radius of cell footprint being $R \propto t^{1/2}$ \citep{Li2013,Sheetz2004,Sheetz2010}. Several mechanistic models of how the cell spreading is achieved after the adhesion to extracellular matrix (ECM) is established \citep{Mahadevan2007,Sheetz2010,Li2013}. However, these papers deal with the characteristic rate of spreading for individual cells. Here, we observe population dynamics directly, observing the stochasticity of cell spreading, and extracting useful information about the underlying kinetics of spreading. This is a useful complementary approach to single cell measurements. We also emphasize that here, and in the rest of this paper, we are discussing isolated cells on a substrate: when cells adhere to each other, their shape transitions are controlled by other mechanisms, based on cadherin and associated pathways~\citep{Buckley2014}.

While reporting and discussing the cell area increase on stiffer substrates, Fig. 5(d) of the paper by Yeung et al.~\citep{Yeung2005} and Fig. 2(A) of the paper by Reinhard-King et al.~\citep{Reinhart2005} also present data on the time-dependence of cell spreading, which already gives a hint for our central experimental finding: the onset of cell spreading does not depend on the substrate. 
In this paper we investigate the time-dependence (kinetics) of the initiation of spreading, asking the question: how long does it take for the cell to recognize the presence of a substrate, and respond by engaging signalling pathways and enacting the required morphological change (spreading on the substrate)? Figure \ref{fig:cells} illustrates the point: plots (a) and (b) show the same cells: immediately after planting on the substrate, and after some time, when several cells have already responded by engaging their spreading. 
We plated two very different cell lines (NIH/3T3 fibroblasts and EA.hy927 endothelial cells) on a variety of substrates that span the range of stiffness from 30 GPa (stiff glass) to 460 Pa (very soft gel), registering the characteristic time at which the initially deposited planktonic cells start to spread. 

We discover three things: [1] the onset of spreading is completely universal, not depending on the stiffness of substrates (in contrast to the final cell morphology, which strongly depends on it); [2] the rate-limiting process, with the characteristic free energy barrier, is the same in both cell lines; [3] the onset of spreading is controlled by a nucleation event, its universal power-law dependence $t^5$ suggesting that there are 5 state changes a newly deposited cell must go through before it is able to spread. We also measure the sum of the free energy changes of these state changes, and find that this, in contrast to the rate-limiting process, depends on the cell line.

\section{Materials and Methods}
\subsection*{Cells and cell culture procedures}
We chose to study endothelial cells and fibroblasts because their adhesion behaviour is important for understanding cardiovascular diseases and tissue engineering.
We used immortalized cell lines: NIH/3T3 murine fibroblasts (obtained from ATCC) and EA.hy927 endothelial cells.

NIH/3T3 fibroblasts are very well characterized, as they have been used in many cell studies since their establishment as cell line; they have also been used in cell adhesion studies, making them a good choice for our experiments~\citep{Todaro1963,Rocha2010}. EA.hy927 is a cell line established in 1983 by the fusion of HUVEC with a lung carcinoma line~\citep{Edgell1983}. It has since become a widely used and thus well characterized cell line, popular in studies of cardiovascular diseases. EA.hy927 cells have also been used for adhesion strength assays~\citep{Han2013}. 

Cells were normally cultured at 37C and 5\% CO2 in Dulbecco's modified Eagle's medium (Greiner) with 10\% fetal bovine serum and 1\% Pen/Strep (solution stabilized, with 10,000 units penicillin and 10 mg streptomycin/mL), from Sigma Aldrich (this standard medium is abbreviated as DMEM). For a comparative study of the role of nutrient in the medium, we also used phosphate-buffered saline (PBS), from Thermo Fisher Scientific during the spreading experiments‎. Cells were subcultured in DMEM every 3 days, at about 70\% confluency, by trypsinization, to avoid the formation of big lumps of cells, thus ensuring that we maintain a single cell suspension. Cells were trypsinized for 5 min (Trypsin-EDTA 0.05\%). The solution was then neutralized by added complete growth medium and centrifuged at 1000 rpm for 5 min. We tested our results on several parallel cell cultures that did not use Pen/Strep, and confirmed that no significant difference was inflicted on our results.

The use of Pen-Strep can be questioned. Antibiotics have been used prophylactically to prevent bacterial infections in cell culture for many years, and they are still being used. It was the introduction of antibiotics that allowed the widespread development of cell culture methods in the first place, as bacterial contamination was a major problem~\citep{Kuhlmann1995}.  However, although toxicity experiments found that antibiotics were harmless to mammalian cells~\citep{Cruickshank1952}, there are concerns about the use of antibiotics in cell culture associated with a neglect of aseptic technique and possible side effects of antibiotics. Many adhesion strength studies use Pen/Strep or other antimycotic or antibiotic solutions in the cell culture, and we followed this procedure as well. We have tested our results on several parallel cell cultures that did not use Pen/Strep, and confirmed that no significant difference was inflicted on our results. 

\subsection*{Substrates of varying stiffness}
To span a wide range of substrate stiffness, we used standard laboratory glass (elastic modulus 30 GPa), and several versions of siloxane elastomers: Sylgard 184 and Sylgard 527, the latter used with the compound/hardener ratio of 1:1 and 5:4. The resulting elastomers were tested on a standard laboratory rheometer (Anton Paar), giving the values of equilibrium modulus $G = 460$ Pa for (Syl527 5:4), $480$ kPa (for Syl184), and $30$ GPa for glass (zero-frequency limit shown in the Supplementary: Fig. S1). For comparison, the stiffness of typical mammalian tissues is commonly reported as: 100 Pa -- 1 kPa in brain tissue; $\sim$3 kPa in adipose tissue; 10 -- 20 kPa in muscle; 30 -- 50 kPa in fibrose tissue; up to a few MPa for bone. We avoided applying the commonly used plasma treatment, as this was making the surface highly uneven on a micron scale, which would affect the adhesion. All surfaces were cleaned by ultrasonication in 96\% ethanol for 15 min, and then incubated with 10 $\mu$g/mL fibronectin in PBS for 45 min.

\subsection*{Experimental procedure and data acquisition}
In our standard cell-spreading experiment, the cell culture dish was inserted into a closed chamber that {maintained controlled temperature with an active water bath, and the $\mathrm{CO}_2$ atmosphere}, while allowing a microscope observation from the top. The cell culture (density $5\times 10^5$ cells per ml, counted by the Neubauer chamber) was placed over the entire substrate. Cells were left to adhere to the substrate for 2 min, at which point the culture dish containing the substrate is filled slowly with fresh medium to reduce the cell density. This was to prevent new cells depositing, and cell clusters forming on the substrate. Only the cells attached to the substrate at this point were included into the subsequent counting. This initial attachment is certainly purely physical, through van der Waals forces and various non-specific cell adhesion molecule head groups. These physically adhered cells, initially spherical in planktonic culture, maintain the high spherical-cap shape with only a small adhesion footprint, as ordinary inflated bilayer vesicles would do as well. This is readily confirmed by the optical interference bands around the cell perimeter, and the lensing effect focusing the light by the short-focal distance near-spherical shape.

After a certain time on substrate, the cells finally engage their specific adhesion-mechanosensing mechanism, and start spreading: to a very widely spread area with highly asymmetric focal adhesions on stiff substrates, or to a round dome-like shape on soft substrates. We are looking to determine the time it takes for the cells to engage this active spreading process. 

	To obtain a population distribution of the onset time of cell spreading, we had to choose a ‘spreading criterion’, which would be clear and easily distinguishable to avoid counting errors. We choose to count the initial onset of visible spreading, seen as the transition between the near-spherical cell initially planted (physically attached) on the substrate, and the cell with adhesion processes engaged and its shape developing an inflection zone around the rim (see Supplementary Fig. S2 for a more detailed illustration and explanation). This morphological transition turns out to be easily identified as the near-spherical cell has a sharp edge, with interference bands in higher magnification, and also a lensing effect of focusing light, which disappears on the transition to a more flattened shape. It must be emphasized, that in order for our cell counting to be meaningful, the cells have to be isolated on the substrate: once the cells come into contact with each other, many other adhesion and mechanosensing mechanisms engage (for example, those based on cadherins), and they spread much more readily and more significantly. That is why our initial cell density was chosen such that the initial attachment is in isolation, and our ‘spreading criterion’ is applied before they spread sufficiently to come in contact (as some cells in Fig. \ref{fig:cells} have done). 

We have carried out many dozens of such spreading experiments, deliberately varying the conditions: comparing cells of different generation age (passage number), medium with and without Penn-Strep, with and without CO2 tent, and at slightly varying pH of the medium -- all on different substrates and at different temperature. The Supplementary Fig. S3 illustrates the robustness and reproducibility of these experiments, which also confirms the meaningful use of the `spreading criterion'.

\begin{figure}
	\centering
	\includegraphics[width=0.43\textwidth]{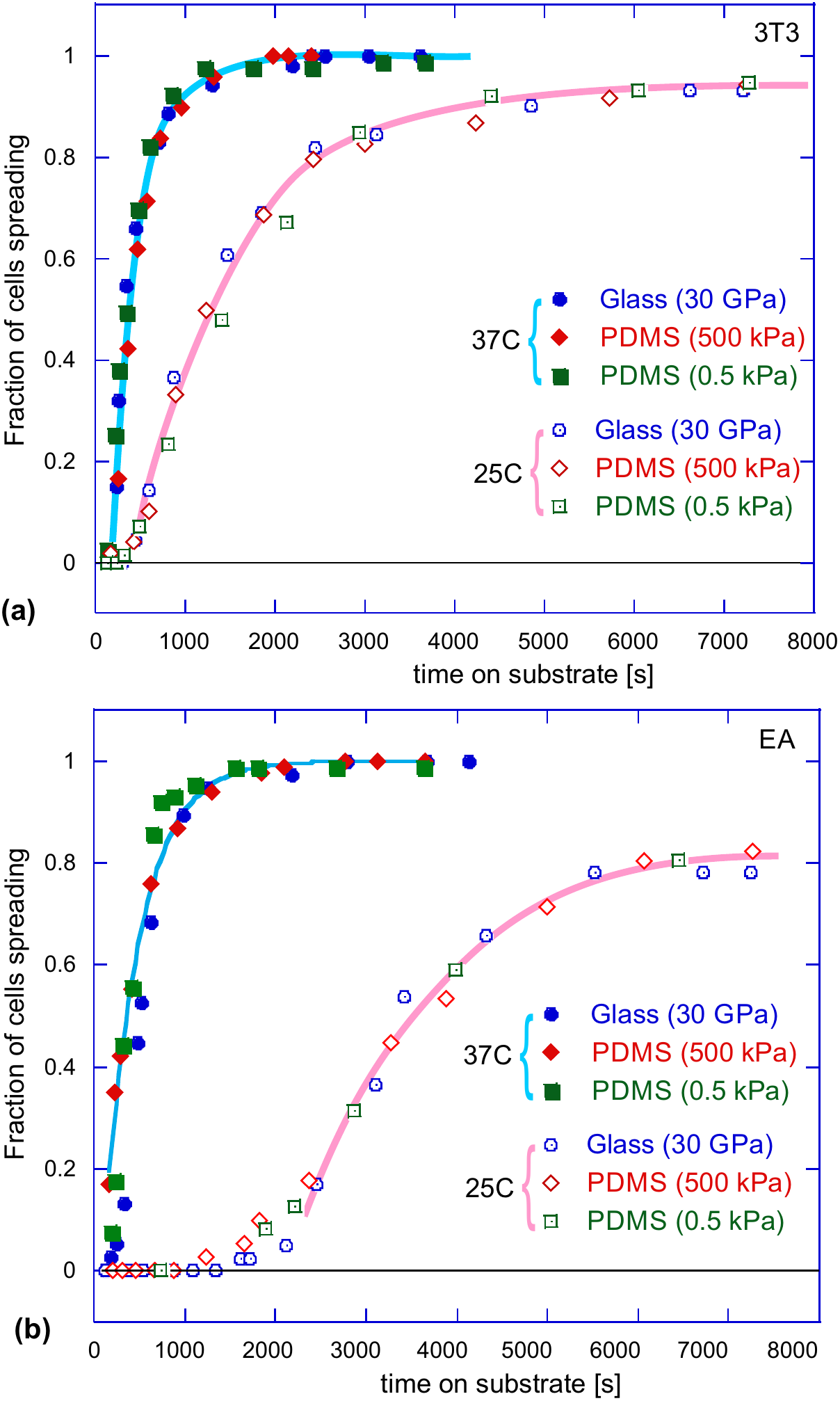}
	\caption{Cumulative population dynamics of cell spreading. Plots (a) and (b) show the growing fraction of cells engaged in spreading on substrates with different stiffness for 3T3 fibroblasts and EA endothelial cells at two different temperatures each. It is clear that the dynamics is not affected by the substrate stiffness, but changes with temperature. In the remainder of this paper, we analyze in detail the long-time behavior of these cumulative curves as they approach saturation, and the behavior at short times when the onset of mechanosensing response occurs.}
	\label{fig:cumuls}
\end{figure}

In each individual experiment (given substrate, fixed temperature, and other parameters), once the cells were deposited on the substrate, and the clock started, we took broad-field microscopic images at regular time intervals, and counted the fraction of cells that have crossed the threshold defined by our spreading criterion -- that is, the sells that have started the active spreading process in response to their mechanosensing cue. This produced a characteristic sigmoidal curve for each experiment (see Fig. \ref{fig:cumuls}): the fraction of cells engaged in spreading starting from zero at $t=0$ and saturating at near-100\% at very long time (if we exclude the occasional cell mortality, which was more of a factor at lower temperatures). The typical sample size was 100-120 cells in each experiment (field of view), however, we have taken many of similar samples and verified the high fidelity of data. The main sources of error were: inconsistency of application of the spreading criterion in image analysis, imperfections of fibronectin coverage on substrate, temperature fluctuations, and of course the natural cell variability. All of these are random errors, with no systematic drift. We were satisfied that the results were reproducible, and errors did not dominate the data trends. The plots in Figs. \ref{fig:cumuls} and \ref{fig:cumuls2} do not include error bars not to obscure distinct data sets, but the reader could gauge this error from Fig. S3 in the Supplementary.

\section{Results} 
We first emphasise that our experiments concurred with the results earlier studies~\citep{Discher2005,Yeung2005,Engler2006}. Cells placed on stiffer substrates spread to larger areas, and were less rounded, for both our cell types. There is also a strong dependence on the ECM protein coverage~\citep{Dubin2004}, but this was not a variable in our study.

The time of initiation of spreading is presented in Fig.~\ref{fig:cumuls}. These two plots shows the fraction of cells that have started spreading at each time after planting on substrate and replacing the medium. The point of steepest gradient in the cumulative curves in marks the most probable time for spreading onset. We see the timing of cell spreading is completely insensitive to the substrate stiffness: the kinetics of spreading response is exactly the same on each substrate.  The work of Sheetz et al.~\citep{Margadant2011} has reported a similar effect (the rate of spreading did not depend on the degree of ECM protein coverage on the surface).   Instead of substrate stiffness, we find the curves in Fig.~\ref{fig:cumuls} are strongly segregated by temperature. 

 \subsection*{Long-time trend: a rate-limiting process.}

\begin{figure}
	\centering
	\includegraphics[width=0.48\textwidth]{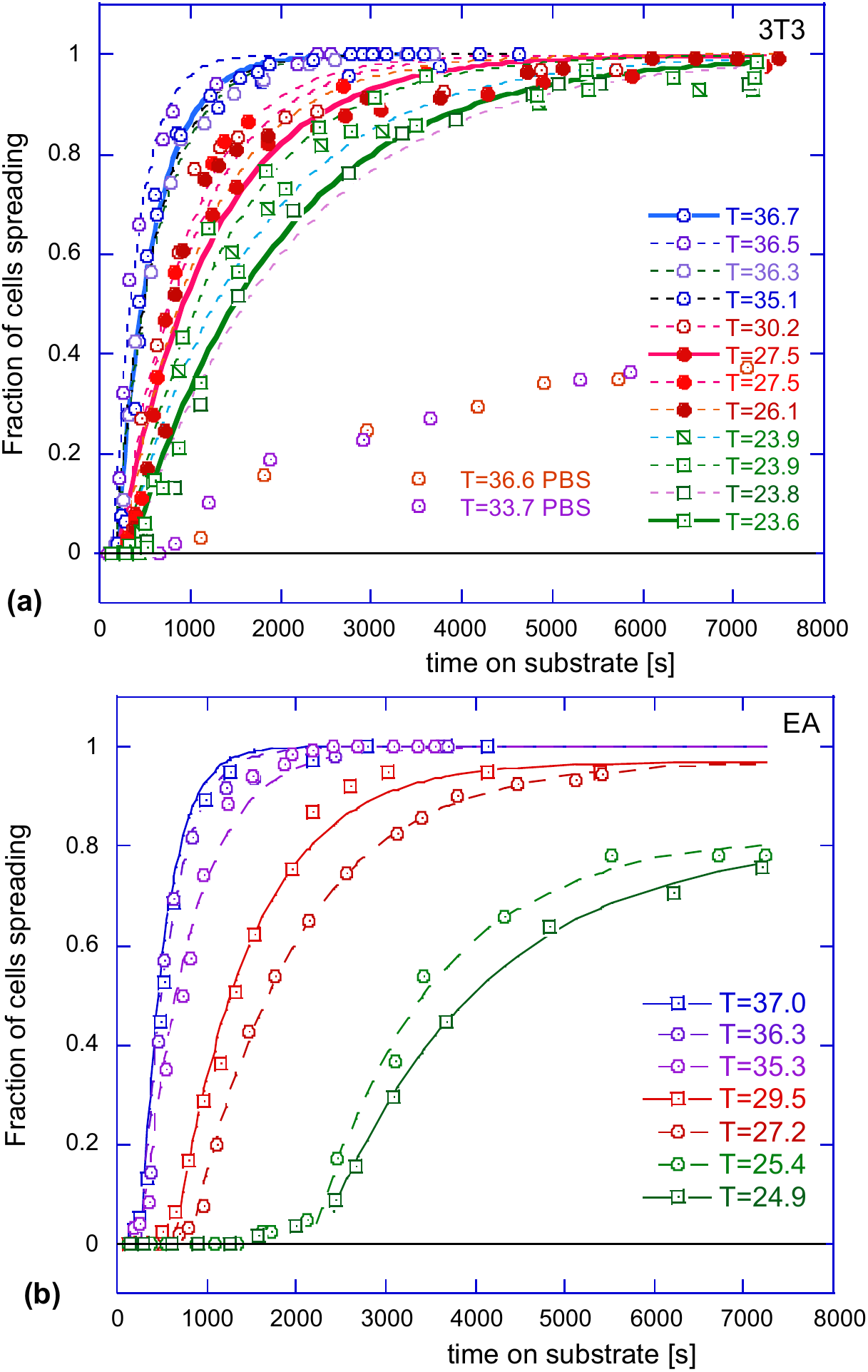}
	\caption{Cumulative population dynamics of cell spreading. Plots (a) and (b) show fraction of spreading cells on glass, at many different temperatures; for 3T3 fibroblasts and EA endothelial cells. Lines in all plots are the fits of the long-time portion of data with the exponential relaxation curves, producing the fitted values of the longest relaxation time $\tau$ (see text).}
	\label{fig:cumuls2}
\end{figure}

To examine the effect of temperature in greater detail, in Fig.~\ref{fig:cumuls2} we plotted the same cumulative spreading fraction curves for the two cell types on glass (as we are now assured that these curves are the same on all substrates). It is noticeable that the initial lag is greater in the EA cells, and that at low temperature the saturation level drops significantly below 100\% -- presumably because more cells disengage (or die) at low temperature, reducing the saturation fraction. The same effect is much enhanced for the the nutrient-starved cells in the PBS medium, see in Fig. \ref{fig:cumuls2}(a): the onset of spreading is very slow in this case, and a large fraction of cells do not engage at all. But the generic sigmoidal shape of the cumulative curve is universal, and the random spread of data within each individual experiment is not excessive. We then look to analyze the trends in this time dependence.

The curves of the generic shape seen in Figs.~\ref{fig:cumuls} and \ref{fig:cumuls2} are encountered in many areas of science, and their characteristic ‘foot’ at early times, especially obvious at lower temperatures, is usually associated with a ‘lag’ in the corresponding process. We will discuss this early-time regime separately, later in the paper, but first we fit exponential relaxation curves to the long-time portion of the data (as the fit lines in Fig. \ref{fig:cumuls2} indicate): $Q(t)=A \cdot (1-\exp[-(t-t_{\mathrm{lag}})/\tau]$. The Supplementary Information gives the table of values of $A$ and $\tau$ for each curve, but it is clear from the plots that the fitting to the single-exponential relaxation law, with just two parameters since $A$ is known for each curve, is very successful. The characteristic relaxation time $\tau$ markedly increases at low temperatures. It is interesting that such a characteristic time associated with the   `spreading of an average cell' has been discussed in \citep{Mahadevan2007}, giving the same order of magnitude (of the order of magnitude 50-100s).

\begin{figure}
	\centering
	\includegraphics[width=0.43\textwidth]{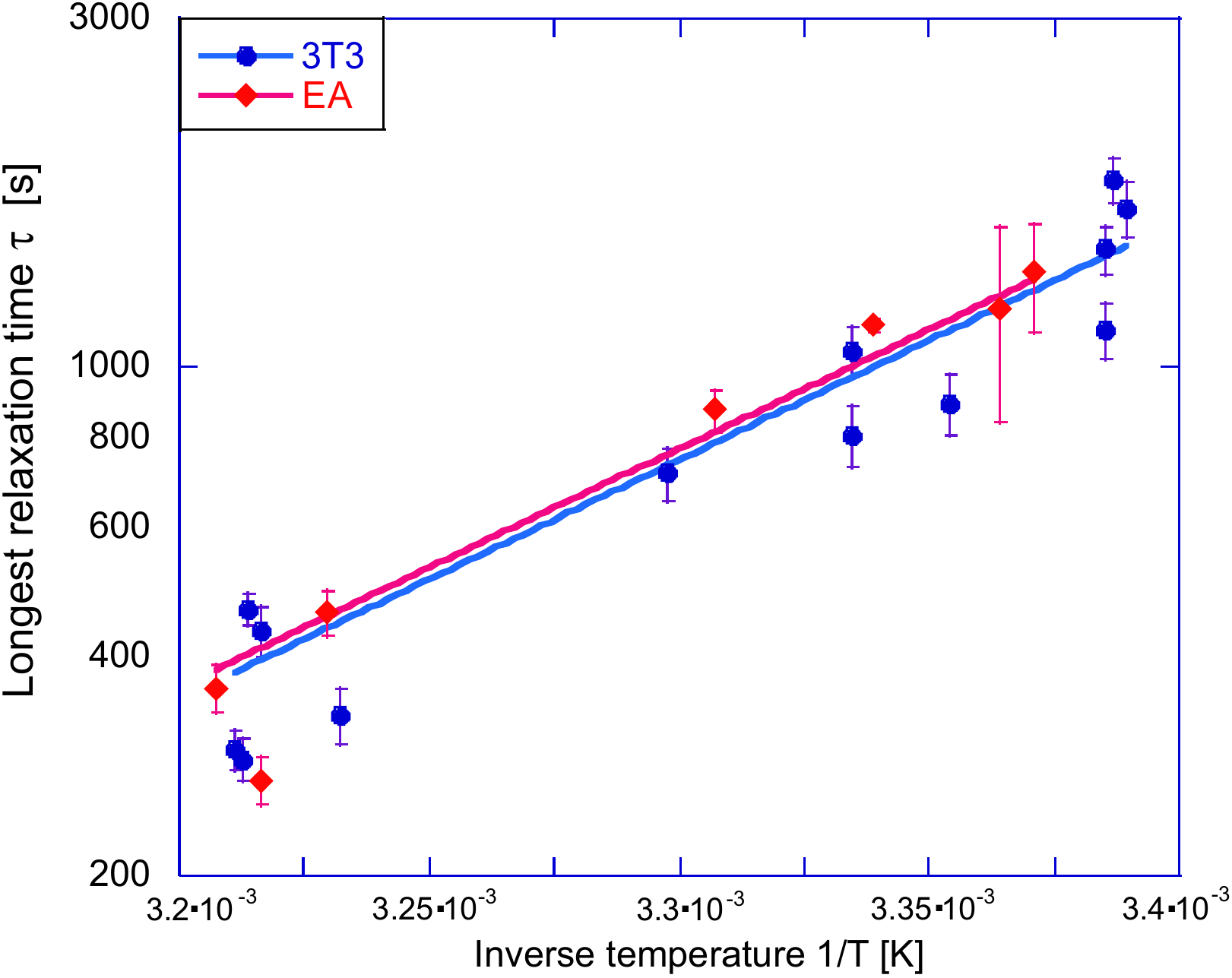}
	\caption{The Arrhenius plot of the longest relaxation time ($\log(\tau)$ vs. inverse absolute temperature) from the exponential fits in Fig.~\ref{fig:cumuls2} (a,b), giving the same value of binding energy $\Delta G\approx 19$ kcal/mol, for both types of cells. }
	\label{fig:arrh}
\end{figure}

To better understand this dependence on temperature, we tested a hypothesis that this relaxation time is determined by the thermally-activated law by producing the characteristic Arrhenius plots of relaxation times, for both cell types, see Fig.~\ref{fig:arrh}. It is remarkable that both cells show almost exactly the same trend of their relaxation time: the rate limiting process in their spreading pathways is the same: $\tau=\tau_0 e^{\Delta G/k_B T}$, with the activation energy $\Delta G\approx 18.7\pm 1.5$ kcal/mol, and the thermal rate of attempts $\tau_0^{-1}\approx 4 \times 10^{10}\mathrm{s}^{-1}$. Both values are very sensible: this magnitude of $\Delta G$ is typical for the non-covalent bonding energy between protein domains~\citep{Zhou2015}, and this rate of thermal collisions is in excellent agreement with the basic Brownian motion values.

\subsection*{Early-time dynamics}

\begin{figure}
	\centering
	\includegraphics[width=0.48\textwidth]{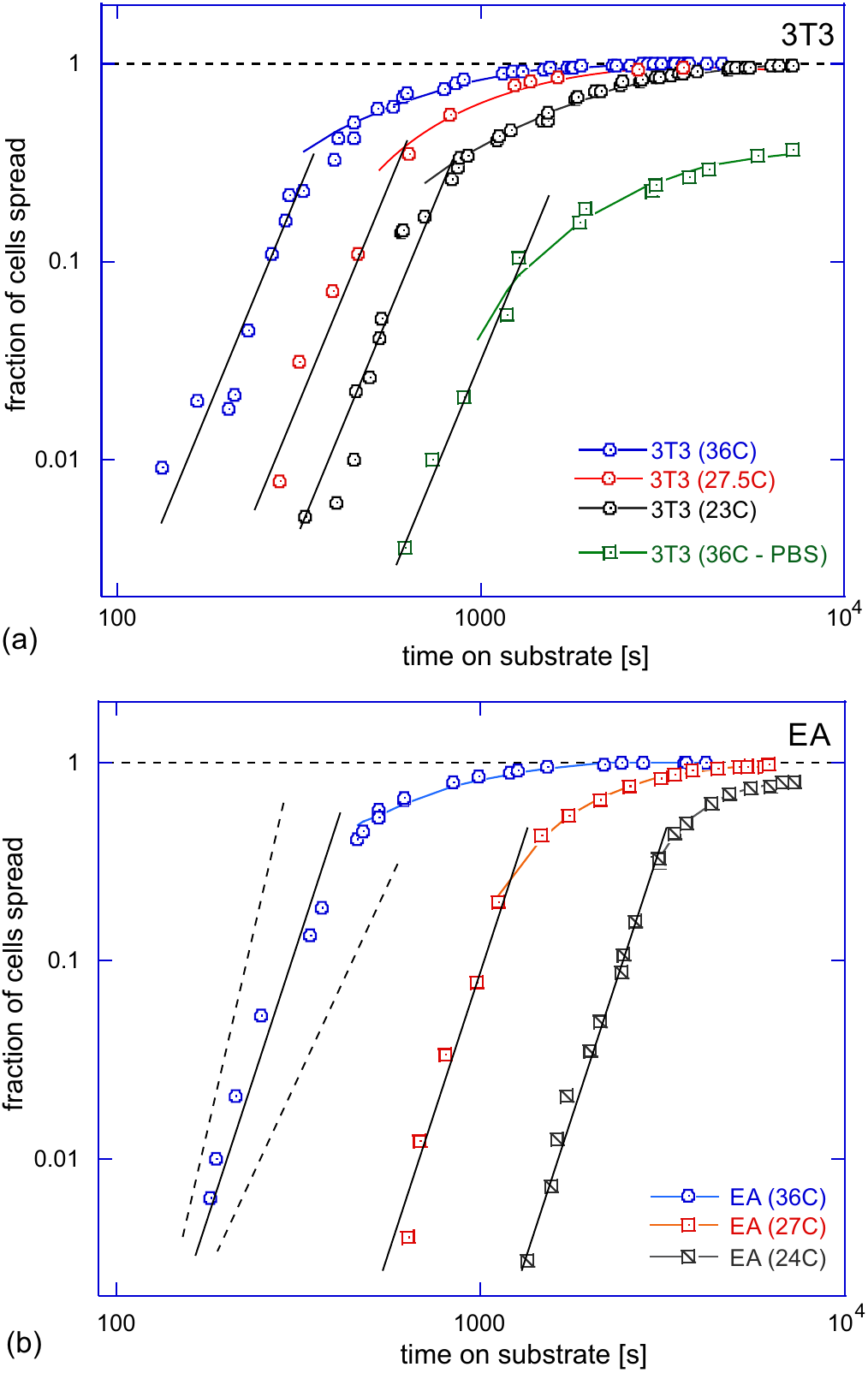}
	\caption{Analysis of the short-time dynamics of cell spreading. Plots (a) and (b) show selected data sets from the Fig.~\ref{fig:cumuls2} (a,b), presented on the log-log scale to enhance the short-time dynamical range. In both plots, the power-law slopes of the short-time data follow the equation: $\alpha t^5$, with the coefficient prefactor $\alpha$ depending both on cell type and on temperature. The dashed line illustrate the slopes of $ t^6$ and $ t^4$ to illustrate the strength of fit.}
	\label{fig:short}
\end{figure}

After discovering that the late-times (rate-limiting) dynamics of the onset of spreading is quite universal across different cells and substrates, it becomes clear that the marked difference between the two cell lines in Fig. \ref{fig:cumuls2} lies in the early-time behavior: something that we have called a `lag' following many similar situations in protein self-assembly. To examine this early-time regime more carefully, we re-plotted the same time series data on the log-log scale in Fig.~\ref{fig:short}. 

This reveals that the process is active from the very beginning ($t=0$) and the plotted value grows as a power-law of time.  The only reason that we appear to see a `lag' is because our experimental technique of counting the cells engaging in spreading did not permit values below 0.01 (1\%) to be resolved in this plot; the same certainly applies to other experimental situations reporting similar kinetic data. The trend illustrated in Fig.~\ref{fig:short} is clear: the early onset of cell spreading follows a universal power law, and the fitting of all our data sets gives $Q(t)=\alpha  t^5$ with very good accuracy, where only the prefactor $\alpha$ depends on temperature and the cell type. We find this result truly remarkable: similarly to the universal value of binding energy that controls thermally-activated rate-limiting relaxation time $\tau$, this very specific $t^5$ power law appears to be the only sensible fit of the early-time data for different cells, temperatures, and substrates.

\begin{figure}
	\centering
	\includegraphics[width=0.48\textwidth]{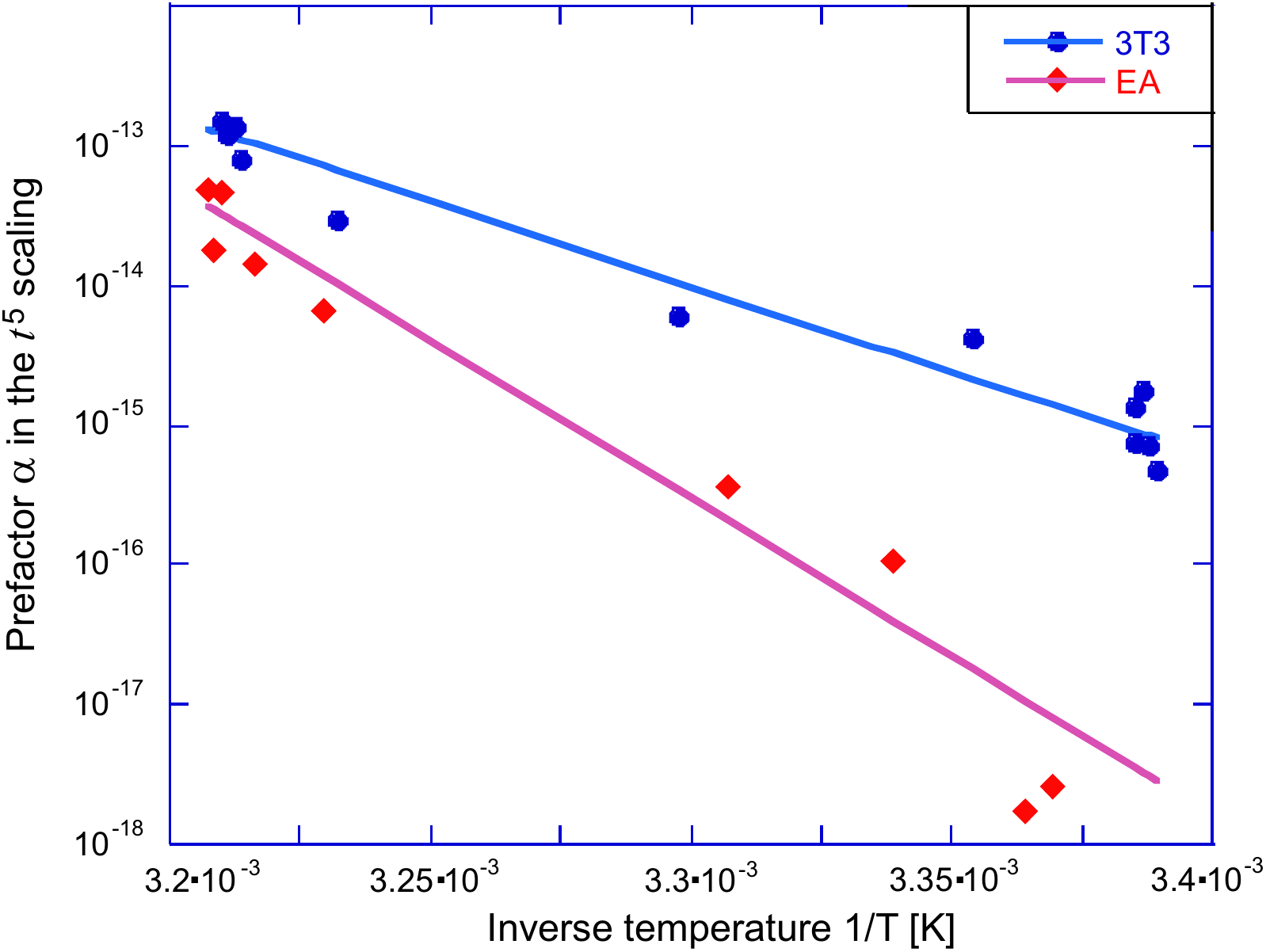}
	\caption{Analysis of the short-time dynamics of cell spreading. The Arrhenius plot of the prefactor $\alpha(T)$, with the fit lines giving the effective activation enthalpy $\Delta H\approx 70$ kcal/mol for 3T3, and 129 kcal/mol for EA. See text, explaining how this value represents the sum of free energy barriers of key proteins assembling into the adhesion complex.}
	\label{fig:short2}
\end{figure}

Again, strong temperature dependence is evident in the subpopulations of cells which start spreading very early: the difference was evident in Figs \ref{fig:cumuls} and \ref{fig:cumuls2}, but is very clearly enhanced in Fig.~\ref{fig:short}. What changes between the data sets  is the prefactor $\alpha$ of the universal power law $\alpha \, t^5$, which has a systematic temperature dependence   (the fitted values of $\alpha(T)$ are listed in the Supplementary table S2). Now expecting the thermally activated behavior, by analogy with the earlier analysis,  we plot these prefactors $\alpha(T)$ on the Arrhenius plot in Fig.~\ref{fig:short2}.  The fitting to $\alpha = \mathrm{const} \cdot e^{-\Delta H/k_B T}$ indeed gives a very reasonable trend, with the activation energies $\Delta H= 70$ kcal/mol for 3T3, and 129 kcal/mol for EA. Note that, in contrast to Fig. \ref{fig:arrh}, here we have a negative exponent, i.e. the parameter $\alpha(T)$ represents a reaction \textit{rate} rather than a relaxation \textit{time}. In the classical Arrhenius-Kramers thermal activation, the process time is shorter as the temperature increases, while the Fig.~\ref{fig:short2} shows the scaling factor $\alpha(T)$  is decreasing as the temperature decreases instead (which is reflected in the overall observation of longer ‘lag time’ in the cumulative curves).
The magnitudes, and the difference in the energy barriers between the two cell types make sense because, due to their biological function, the mechanosensing process in fibroblasts should start faster. However, we have so many different quantitative facts and trends that it is necessary to look much more carefully at what we understand about mechanosensing.

\section{Discussion}

In classical physics, early-time power law kinetics are a hallmark of self-assembly processes such as polymerization or aggregation~\citep{Hofrichter1974}. In this case, we are looking at a process of self-assembly within the cell. The exponent of the power law gives us some idea of how many important assembly steps there are. But, what exactly are we assembling? To us, it seems likely that we must be observing the formation of adhesion points and complexes that allow the cell to bind onto its ECM environment and begin spreading.

It is already well-known that disruption of the integrin-fibronectin linkage completely halts cell spreading~\citep{Zhang2008,Price1998}. Integrins are transmembrane receptors (references) linking the cell to the matrix in focal adhesions \citep{Hynes2002,Giancotti2000,Guan1991}. To attach to their ligands, they need to be activated \citep{Kim2011,Shattil2010}. However, in isolation, integrin pairs will lie in their inactive state, unable to bind to fibronectin (or other ECM proteins containing the RGD motif).

Much of the literature on focal adhesions sees the attachemnt of the talin head domain to integrin tails as an important activation step~\citep{Tadokoro2003,Wegener2007,Moser2009}. Talin is a key protein in mature and nascent adhesions, linking integrins to the actin cytoskeleton, and providing a scaffold for other focal adhesion proteins (see, for example~\citep{Geiger2009}). For the onset of spreading, there is some conflict in the literature: in the study by Zhang et al.~\citep{Zhang2008}, where they confirmed that integrin linkage was essential to the onset of spreading, they actually depleted both types of talin, and found that the onset of spreading was not fully inhibited, although spreading was severely limited. This could indicate that talin was not needed for the activation of integrins during the onset of spreading. However, a subsequent knock-out study of talin (among other proteins)~\citep{Theodosiou2016} found that spreading was actually completely inhibited by the removal of talin (although partial function was restored by the addition of Mn\textsuperscript{2+}). In that work, the authors note that the experimental methods (si-RNA transfection) employed in previous studies left residual amounts of proteins in the cell, and that there may well have been enough talin left in depleted cells to form nascent adhesions. Indeed, in their paper, Zhang et al. say that the decrease in talin2 levels (talin1 was not expressed in their cell lines) was between 40-68\%.

In fact, Theodosiou et al. implicate three further players: kindlins, paxillin, and focal adhesion kinase (FAK)~\citep{Theodosiou2016}. This is not a new finding, or point of view: since the early discovery of the key role of FAK in the integrin adhesome \citep{Guan1991,Sieg2000,Parsons2003}, it was understood that is is the FAK activation that produces the chemical cue for the subsequent cell mechanosensing pathways via Src, Rho, Rac and Cdc42, as well as Erk~\citep{Huveneers2009,Price1998,Schwartz2000,Welch2015}. Theodosiou et al. found that chemical inhibition of FAK reduced lamellopodia formation in cells to the level of kindlin knock-out cells~\citep{Theodosiou2016}. The formation of these lamellopodia and the initiation of isotropic cell spreading was therefore found to be dependent on FAK activation. A recent model of FAK as a mechanosensor~\citep{Bell2017} shows how the rate of its activation is sensitive to the stiffness of substrate, and the cytoskeletal pulling force. Importantly, when the force is low (as we would expect at early times, before the mechanosensing pathways are activated and the cytoskeletal forces increase), this rate is controlled only by the bonding energy between its FERM and kinase domains, not the stiffness.

\begin{figure*}
	\centering
    \includegraphics[width=0.7\textwidth]{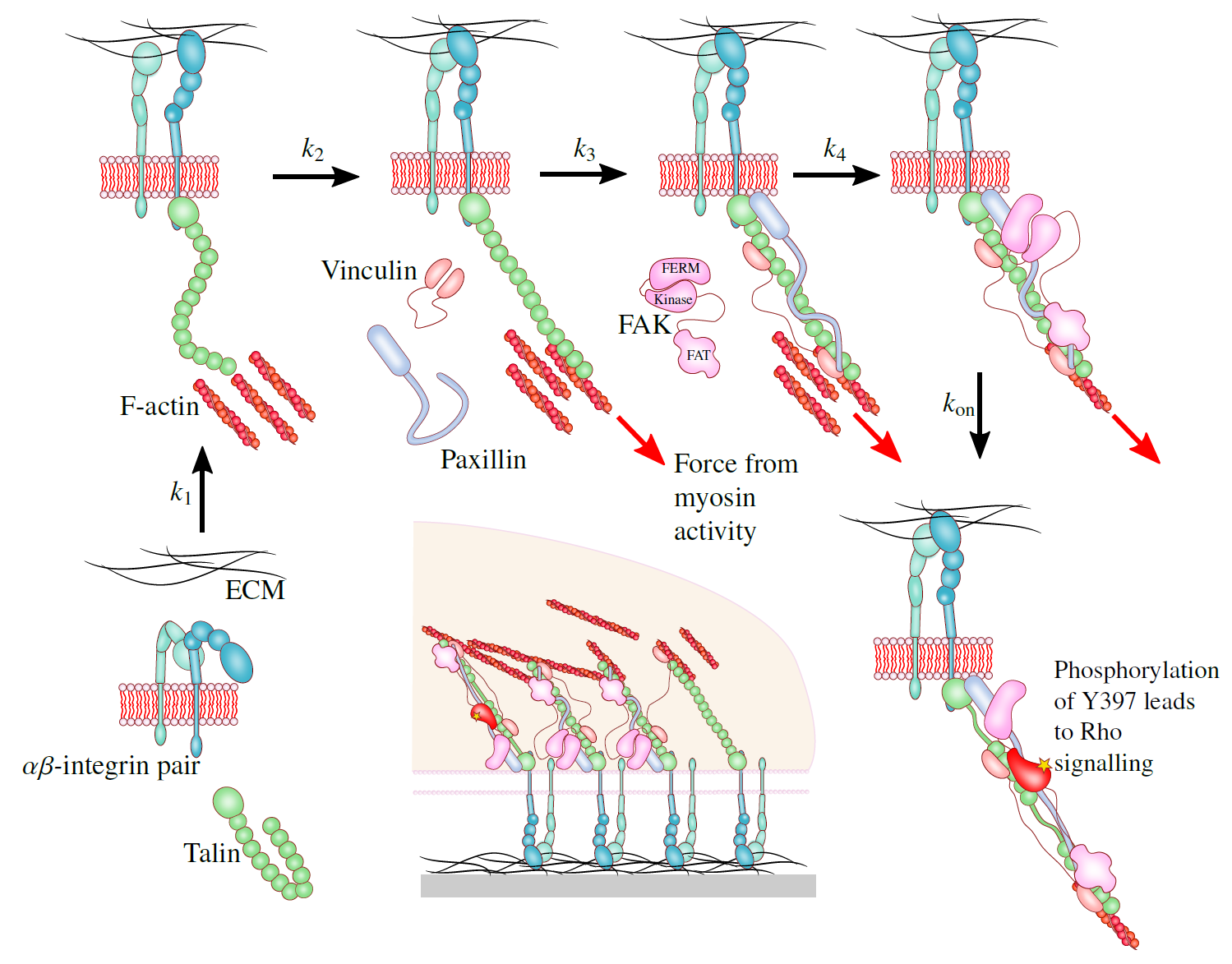}
	\caption{A possible assembly sequence of a mechanosensor complex. Our analysis suggests that there are five distinct ‘slow’ stages illustrated in the sequence, with their respective rates $k_1$-$k_4$ and the rate of FAK activation $k_\mathrm{on}$ (controlled by the free energy barrier $\Delta G\approx 19$ kcal/mol, cf. Fig. \ref{fig:arrh}). The product of the five rate constants $\alpha=k_1 k_2 k_3 k_4 k_\mathrm{on}$ is what we measure in the Arrhenius plot Fig. \ref{fig:short2}. In the center is a sketch of forming focal adhesion cluster, where the individual mechanosensor complexes in various stages of development/turnover are bound by vinculin and actin crosslinking (see text for details and references).}
	\label{fig:scheme}
\end{figure*}

FAK clearly sits at the centre of the adhesion signaling network~\citep{Zaidel2007}. But the minimal composition of the whole adhesion-mechanosensing complex in the nascent adhesions, as well as the rate of its assembly and turnover, remain a question of active research and debate. Kindlins are known to be a necessary partner for talin in integrin activation~\citep{Geiger2009,Kim2011,Moser2009}. The F3 subdomain of a FERM domain mediates an interaction with $\beta$-integrin tails, and `cooperates' with the talin head domain in integrin activation~\citep{Moser2008}. 
Paxillin is another player in the adhesion network~\citep{Geiger2009,Theodosiou2016,Zaidel2007}. In particular, in the nascent adhesions formed at the onset of spreading, kindlin was directly binding paxillin; paxillin was then recruiting FAK to these nascent adhesions. On the other hand, the important role of vinculin in several processes in the integrin-talin-FAK adhesion complex appears to be relevant mostly at the mature focal adhesion stage \citep{Hemmings1996,Margadant2011,Yao2014}, and we believe its role is to bind different adhesion complexes into a dense focal adhesion raft.

How does this information tie in with our results? A recent molecular-dynamics simulation~\citep{Zhou2015} has explicitly calculated the bonding energy between FERM and Kinase domains of FAK as $\Delta G\approx 17$ kcal/mol. Breaking this bond is the essential step of FAK activation. If we associate this barrier with the longest relaxation time examined in Fig.~\ref{fig:arrh}, the agreement of the  $\Delta G$ values is remarkably close. According to reaction rate theory, this energy barrier is the largest one of the assembly process, as it produces the long-time `bottleneck' in the population dynamics of the onset of spreading.  

According to our analysis, the cell must undergo 5 changes of state before it can initiate the spreading response, with the last being the FAK activation process \citep{Bell2017}, see Supplementary Information for detail. The possible candidates for the other 4 reaction steps must have a rate slow enough to be counted in the first data points, see Fig. \ref{fig:scheme} for an illustration. Images of cells were taken approximately every minute, and so it is impossible to resolve fast processes with rates of $k>1\mathrm{min}^{-1}$ using our data. For instance, the binding of integrins to fibronectin does not fit this criterion: it has been seen that the binding of integrins to an antibody ligand in the presence of different cations has a characteristic binding time of $0.01-1$ms~\citep{Hu1996}; this is much faster than we could resolve in our experimental data. In order to form the force-bearing chain from integrin to F-actin of cytoskeleton, we see the following reactions necessary: [a] 
 the binding of talin and kindlin to integrins, [b] the binding of paxillin to kindlin, [c] the binding of talin to F-actin, [d] the binding of FERM domain of FAK to talin, [e] the binding of FAT domain of FAK to paxillin, and [f] the binding of FAK/paxillin to the F-actin. 
It is difficult to find any estimates of the rates of these processes. One can find evidence for the fast strengthening of focal adhesions under load~\citep{Strohmeyer2017}, but this is not the same as the assembly of these complexes at the onset of spreading. 
Our experiments suggest that 4 of these reactions are quite slow (accounting for the need of protein localization on the complex); we cannot be certain which, but we have measured the combined activation energy of these four reactions (Fig. \ref{fig:short2}) in 3T3 and EA cells. Only once the full force-chain of the integrin adhesome is assembled, the mechanosensor produces the signal for the cell to modify its morphology to the substrate. 

The unusual feature of this work is the use of population dynamics of spreading cells to infer details of the microscopic processes governing the cell response to an external substrate. By linking the results to nucleation theory, details of which are given in Supplementary Information, we found a novel way of looking at the onset of cell spreading as a problem of complex assembly.

\subsection*{Author contribution}
ALR carried out all experiments. SB, ALR and EMT carried out different elements of data analysis. SB and EMT wrote the paper.

\subsection*{Acknowledgements}
The authors acknowledge many helpful discussions with K. Franze, K. Chalut, A. B. Kolomeisky, and H. Welch. The comments of U. S. Schwarz, and the experimental support in the cell culture lab by E. Nugent and F. Morgan are much appreciated. This work has been funded by EPSRC (grants EP/M508007/1 and EP/J017639), and the Ernest Oppenheimer Trust in Cambridge.

\end{document}